\documentclass[dvips]{article}
\usepackage{graphicx}
\newcommand{\doublespacing}{\let\CS=\@currsize\renewcommand{\baselinesstrech}
{2.0}\tiny\CS}

\begin{document}

\textwidth 16cm
\newcommand{\bd}{\begin{document}}
\newcommand{\ed}{\end{document}}
\newcommand{\bc}{\begin{center}}
\newcommand{\ec}{\end{center}}
\newcommand{\bfr}{\begin{flushright}}
\newcommand{\efr}{\end{flushright}}
\newcommand{\lt}{\left}
\newcommand{\rt}{\right}
\newcommand{\vs}{\vspace}
\newcommand{\hs}{\hspace}
\newcommand{\beq}{\begin{equation}}
\newcommand{\eeq}{\end{equation}}
\newcommand{\lb}{\linebreak}
\newcommand{\pb}{\pagebreak}
\newcommand{\mb}{\makebox}
\newcommand{\fb}{\framebox}
\newcommand{\mc}{\multicolumn}
\newcommand{\ben}{\begin{enumerate}}
\newcommand{\een}{\end{enumerate}}
\newcommand{\bit}{\begin{itemize}}
\newcommand{\eit}{\end{itemize}}
\newcommand{\ol}{\overline}
\newcommand{\un}{\underline}
\newcommand{\lefq}{\lefteqn}
\newcommand{\ba}{\begin{array}}
\newcommand{\ea}{\end{array}}
\newcommand{\beqa}{\begin{eqnarray}}
\newcommand{\eeqa}{\end{eqnarray}}
\newcommand{\beqas}{\begin{eqnarray*}}
\newcommand{\eeqas}{\end{eqnarray*}}
\newcommand{\bfg}{\begin{figure}}
\newcommand{\efg}{\end{figure}}
\newcommand{\bds}{\begin{displaymath}}
\newcommand{\eds}{\end{displaymath}}
\newcommand{\btb}{\begin{tabbing}}
\newcommand{\etb}{\end{tabbing}}
\newcommand{\para}{\parallel}
\newcommand{\pad}{\partial}
\newcommand{\nn}{\nonumber}
\newcommand{\la}{\leftarrow}
\newcommand{\ra}{\rightarrow}
\newcommand{\lgla}{\longleftarrow}
\newcommand{\lgra}{\longrightarrow}
\newcommand{\La}{\Leftarrow}\newcommand{\Ra}{\Rightarrow}
\newcommand{\Lra}{\Leftrightarrow}
\newcommand{\Lgla}{\Longleftarrow}
\newcommand{\Lgra}{\Longrightarrow}
\newcommand{\bm}{\boldmath}
\newcommand{\lan}{\langle}
\newcommand{\ran}{\rangle}
\renewcommand{\a}{\alpha}
\renewcommand{\b}{\beta}
\newcommand{\g}{\gamma}
\newcommand{\G}{\Gamma}
\renewcommand{\d}{\delta}
\newcommand{\eps}{\epsilon}
\newcommand{\Th}{\Theta}
\newcommand{\s}{\sigma}
\newcommand{\lam}{\lambda}
\newcommand{\D}{\Delta}
\newcommand{\vare}{\varepsilon}
\newcommand{\pr}{\prime}
\newcommand{\ro}{\rho}
\newcommand{\nab}{\nabla}
\newcommand{\m}{\mu}
\newcommand{\n}{\nu}
\newcommand{\Sg}{\Sigma}
\newcommand{\p}{\pi}
\newcommand{\R}{I\!\!R}
\newcommand{\om}{\omega}
\newcommand{\Om}{\Omega}
\newcommand{\ze}{\zeta}
\newcommand{\vart}{\vartheta}
\newcommand{\tri}{\triangle}
\newcommand{\f}{\frac}
\newcommand{\iny}{\infty}
\newcommand{\pro}{\propto}
\bc
{\Large \bf New Exactly Solvable Isospectral Partners for ${\cal PT}$ Symmetric Potentials}
\ec

\vs{1cm}

\bc
{\it Anjana Sinha{\footnote {e-mail : anjana23@rediffmail.com}}\\
Department of Applied Mathematics \\
Calcutta University \\
92, APC Road, Kolkata - 700 009}
\ec

\vs{.25cm}

\bc
{and}
\ec

\vs{.25cm}

\bc
{\it Pinaki Roy{\footnote {e-mail : pinaki@isical.ac.in}} \\
Physics \& Applied Mathematics Unit \\
Indian Statistical Institute \\
Kolkata - 700 108}
\ec

\vs{1cm}

\vs{1cm}

\bc
{\large {\un{Abstract}}}
\ec

We examine in detail the possibilty of applying Darboux transformation to non Hermitian hamiltonians. In particular we propose a simple method of constructing exactly solvable ${\cal{PT}}$ symmetric potentials by applying Darboux transformation to higher
 states of an exactly solvable ${\cal{PT}}$ symmetric potential. It is shown 
that the resulting hamiltonian and the original one are pseudo supersymmetric partners. We also discuss application of Darboux transformation to hamiltonians with spontaneously
broken ${\cal{PT}}$ symmetry.
\pb

\section{Introduction}

Ever since it was conjectured by Bender et al that 
some non-Hermitian Hamiltonians exhibiting
symmetry under the combined transformation of {\it parity} 
($ {\cal{P}}$ : $ x \ra -x $), and {\it time reversal} ($ {\cal{T}} $ : 
$ i \ra -i $) admit real eigenvalues \cite{ben,mei}, non-Hermitian 
Hamiltonians have been the basis of many recent works on 
${\cal{PT}}$ symmetry and pseudo-Hermiticity \cite{znojil, cannata}, because of 
intrinsic interest and their possible applications 
in molecular physics, quantum chemistry, superconductivity,
quantum field theory and others.
 
On the other hand there are not many examples of exactly solvable complex
potentials (both ${\cal{PT}}$ invariant as well as
otherwise). However, as in the Hermitian case, there have been attempts 
to expand the class of exactly solvable non Hermitian potentials by using
different methods \cite{ikonic,milan,rkrc}. In the Hermitian case a popular method 
to obtain new exactly solvable potentials is to apply Darboux transformation
\cite{mat} to the ground state of an exactly solvable potential. However,
when applied to the excited states, this transformation produces not just one
isospectral potential, but a number (depending on the nodes of the wave
function) of nearly isospectral potentials which are defined not over the
whole domain, but in disjoint intervals \cite{robnik}.
Here our objective is to apply Darboux transformation to non Hermitian potentials and it will be shown that for such potentials, it is possible to have wave
functions without nodes on the real line, by a reasonable choice of
parameters. In the present article, we shall use this result to construct non
trivial isospectral partners of exactly solvable complex potentials. In
particular, we shall apply the Darboux transformation 
to the well known ${\cal{PT}}$ symmetric Scarf II potential
\beq
V(x) = - V_1 sech ^2 x - i V_2 sech ~x ~ tanh ~x, ~~~~~~~~~~~~~~~ V_1 > 0, ~
V_2 \neq 0 \label{scarf2}
\eeq
and generate a series of new exactly solvable non-Hermitian potentials with real spectrum. 

We note that in the case of Hermitian quantum mechanics Darboux transformation is equivalent to supersymmetry \cite{khare1}. However this is not so in the non Hermitian case. So it is natural to ask whether there exists any symmetry which relates the two Hamiltonians i.e., the original and the one obtained by Darboux transformation. The answer to this question is in the affirmative and it will be shown that the two Hamiltonians are related by pseudo supersymmetry \cite{mostafa}. In other words the hamiltonians obtained by intertwining are pseudo supersymmetric partners.

Finally we shall examine the problem of applying Darboux transformation to models with spontaneously broken ${\cal{PT}}$symmetry. It is known \cite{zafar} that models with spontaneously broken ${\cal{PT}}$symmetry exhibit a complex spectrum and all the energy levels appear as complex conjugate pairs. It will be shown that if Darboux transformation is applied to such a system one gets a potential with complex energy eigenvalues but as singlets. 

The organisation of the paper is as follows : in section 2 we briefly present the Darboux construction; in section 3 we construct new ${\cal{PT}}$ symmetric potentials; in section 4 we show that the partner Hamiltonians are connected by pseudo supersymmetry; in section 5 we examine the nature of the spectrum obtained by applying Darboux transformation to a potential with spontaneously broken ${\cal{PT}}$ symmetry and finally section 6 is devoted to a discussion.
\section{Darboux Transformation}

To make the article self contained we start with a brief review of the Darboux
transformation \cite{mat, robnik}. A particle moving in the potential 
$v(x)$ (real or complex) is characterised by the Hamiltonian
\beq
H = - \f{d^2}{dx^2} + v (x) \label{hamiltonian}
\eeq
(The units used are $ \hbar = 2m = 1$ for convenience). \\
If the particle is in the $m^{th}$ state, (i.e., $m$ is the quantum number
equal to the number of nodes of the $m^{th}$
eigenfunction $ \psi _m (x) $ of the
starting potential $v (x)$), and the energy scale is adjusted so that the
$m^{th}$ energy eigenvalue is exactly zero ($ E_m  = 0 $), then the
Schr\"{o}dinger equation reads
\beq
H \psi _m  = \lt( - \frac{d^2}{dx^2} + v (x) \rt) \psi _m  = 0 \label{schro}
\eeq
Equation (\ref{schro}) has a potential
\beq
v (x) = \frac{\psi _m ^{\pr \pr}}{ \psi _m} \label{pot}
\eeq 
which is regular everywhere, so that the Hamiltonian in (\ref{schro}) may be 
represented as
\beq
H  = \lt( - \frac{d^2}{dx^2} + \frac{\psi _m ^{\pr \pr}}{ \psi _m} \rt) 
\eeq
Thus if the general solution $ \psi = \psi (x) $ 
of the Schr\"{o}dinger equation 
\beq
\f{d^2 \psi }{dx ^2} + \lt[ \eps - v(x) \rt] \psi = 0 
\eeq
is known for all values of $ \eps $, and for a particular value of $ \eps =
E_m $, the particular solution is $ \psi _m $, then the general
solution of the equation 
\beq
\f{d^2 \phi  }{dx ^2} + \lt[ E - u(x) \rt] \phi  = 0 
\eeq
with
\beq
\ba {lcl}
\displaystyle u(x) 
&=& \displaystyle \psi _m (x) \f{d^2}{dx^2} \lt( \f{1}{ \psi _m (x) } \rt) \\
&=&  \displaystyle 2 \lt( \f{ \psi _m ^{\pr} }{ \psi _m } \rt) ^2 - 
\lt( \f{ \psi _m ^{\pr \pr} }{ \psi _m } \rt) \label{partner}
\ea
\eeq
\beq
E = \eps - E_m
\eeq
for $ E \neq 0 $ is
\beq
\ba {lcl}
\displaystyle \phi _n  (x) 
&=&  \displaystyle \psi _m (x) \lt\{ \f{ \psi _n (x) }{ \psi _m (x) } \rt\} ^{\pr} \\
&=&  \displaystyle \psi _n ^{\pr} (x) 
- \lt( \f{ \psi _m ^{\pr} (x) }{ \psi _m (x) } \rt) \psi _n (x)\label{wavepartner} 
\ea
\eeq
Choosing different $\psi _m$, one obtains a series of non-trivial partners 
$u(x)$ given by (\ref{partner}), which contain all the states of the original potential $v(x)$ except the $m^{th}$ state, i.e. the one corresponding to the eigenstate $\psi _m$. 
It is easy to observe that the partners $u(x)$ and $v(x)$ are related by 
\beq
v(x) = W_m (x) ^2 - W_m ^{\pr} (x)
\eeq
\beq
u(x) = W_m (x) ^2 + W_m ^{\pr} (x)
\eeq
where 
\beq
W_m (x) = \lt( - ~\f{\psi _m ^{~\pr}}{\psi _m} \rt)  \label{superpotential}
\eeq
Though $ \psi _m ^{\pr} \neq 0 $ 
at the nodes $x_j$, $j=1,2,3, \cdots $ of $ \psi _m $,  $W _m (x) $ 
has singularities at these points. However, since the second derivative 
$ \psi _m ^{\pr \pr}$ also vanishes at the nodes, from (\ref{pot}) 
$v(x)$ is regular everywhere.
In case of Hermitian SUSY QM, $ W_m $ is the superpotential, and 
$ W_m ^2(x) \pm W_m ^{\pr} (x)$ are called SUSY-$m$ partner potentials
\cite{robnik}. However, one can 
construct the partners for $m=0$ only, as for non-zero $m$, $W_m$ becomes singular and as a consequence such potentials are not defined over $ R $ but over disjoint intervals, the 
number of intervals depending on the value of $m$. Thus,
for $m=1$, there are two potentials, each of them defined on a
semi-infinite domain, for $m=2$ there is one potential on a
finite domain between nodes $x_1$ and $x_2$, and two potentials 
on the two semi-infinite domains $(- \infty, x_1]$ and 
$[x_2, + \infty )$, and so on \cite{robnik}. 
In the case of non-Hermitian quantum mecahnics, 
$W_m (x)$ is no longer singular (and it is not the {\it superpotential} anymore).
As a consequence the 
new potentials do not have singularities on the real axis and are defined on $(-\infty,\infty)$.
Thus, if one of the partner potentials is exactly solvable, this 
formalism enables one to construct an infinite number of 
exactly solvable, non-trivial partners defined on the entire real line
$(- \infty, + \infty)$, unlike in the case of Hermitian quantum mechanics.

\vs{1cm}

\section{New exactly solvable ${\cal{PT}}$ symmetric potentials}

In this section we shall construct isospectral partners of the complexified Scarf II potential. This potential is given by 
\beq
V(x) = - V_1 sech ^2 x - i V_2 sech ~x ~ tanh ~x, ~~~~~~~~~~~~~~~ V_1 > 0, ~
V_2 \neq 0 
\eeq
and it has been studied by various authors as it is not only invariant under
${\cal{PT}}$ symmetry, but also ${\cal{P}}$-pseudo Hermitian. 
This exactly solvable model has certain interesting properties. 
It has a discrete spectrum that admits 
both real as well as complex conjugate energies, 
depending on the relative strengths of its parameters $V_1$ and $V_2$. 
The normalized wave functions for this potential are well known, being 
given by \cite{zafar}
\beq
\psi _n (x) = \f{ \Gamma \lt( n - 2p + \f{1}{2} \rt) }{ n! \Gamma \lt(
\f{1}{2} - 2p \rt)}~ z^{-p}~(z^*)^{-q} ~ P_n ^{-2p - \f{1}{2}, ~ -2q -
\f{1}{2}} ( i~sinh ~x) \label{wavescarf}
\eeq
where $P_n ^{\alpha, \beta}$ are the Jacobi polynomials given by
\cite{handbook} 
\beq
P_n ^{\alpha,\beta} (i~sinh ~x) = \f{\Gamma (n+ \a +1)}{\Gamma (n+1)
\Gamma (\a +1)} ~ 
F \lt( -n, n + \alpha + \beta + 1; \alpha + 1; z \rt)
\eeq
and
\beq
z = \f{1-i~sinh~x}{2}
\eeq
\beq
p = - \f{1}{4} \pm \f{1}{2} \sqrt{\f{1}{4} + V_1 + V_2 } = - \f{1}{4} \pm \label{p}
\f{t}{2} 
\eeq
\beq
q = - \f{1}{4} \pm \f{1}{2} \sqrt{\f{1}{4} + V_1 - V_2 } =  - \f{1}{4}
\pm \f{s}{2}\label{q}
\eeq
$t$ and $s$ are defined with only the positive sign in the discriminant in $p$
and $q$.
The energy eigenvalues are obtained as 
\beq
E_n = - \lt( n -p - q  \rt)^2, ~ ~ ~ ~ ~  n=0, 1, 2,... < 
\lt( \f{s+t-1}{2} \rt)
\eeq
Since $V_1 > 0 $, two cases arise for real $V_2$, depending on the 
relative strengths of the
real and imaginary parts of the potential : 

\vs{.5cm}

\noindent
1. ~~ For $ |V_2| \leq V_1 + \f{1}{4} $ \\
In this case the potential and the wave functions are $\cal{PT}$ invariant, $p$ and $q$ are real  and one gets a real bound state spectrum.
In addition to the potential (1), the wave functions $\psi _n (x)$ 
given in (15) are also 
${\cal{PT}}$ invariant. Note that due to 
normalisation requirements only the values with the positive 
sign are allowed in (\ref{p}) and (\ref{q}). 

\vs{.5cm}

\noindent
2. ~~ For $ |V_2| > V_1 + \f{1}{4} $ \\
${\cal{PT}}$ symmetry is spontaneously broken, as though the potential is
${\cal{PT}}$ invariant, the wave functions are no longer so. 
Either $p$ or $q$ is complex, 
and all energies occur as complex conjugate pairs. Real energies are
conspicuous by their absence. 

\vs{.5cm}

\noindent
For purely imaginary $V_2$, however, the potential (\ref{scarf2}) 
is real, possessing only real energies.

\vs{.5cm}

\noindent
However, when $V_2 $ has both real and imaginary parts, the potential (\ref{scarf2}) loses its
${\cal{PT}}$ invariance, and so will not be considered in this study.

\vs{.5cm}

We now consider the first case when ${\cal{PT}}$ symmetry is unbroken. Now, using the explicit solution for the normalized wave function 
(\ref{wavescarf}), 
we obtain 
\beq
\ba {lcl}
\displaystyle W _m (x) 
&=&  \displaystyle - \f{ \psi _m ^{\pr} (x) }{ \psi _m (x) } \\
&=&  \displaystyle (p+q) tanh ~x - i(p-q) sech ~x 
+ \f{mb}{c} \f{ F \lt( -m+1, b+1; c+1; z \rt) }{ F \lt( -m, b; c; z \rt) }\label{super}
\ea
\eeq 
where $b$ and $c$ stand for 
\beq
b = -2p -2q +2
\eeq
\beq
c = -2p + \f{1}{2}
\eeq
The exactly solvable potential $ U ^{(m)}(x)$, which is isospectral to the 
Scarf II potential (except for the $m^{th}$ state),
is obtained from the formula
\beq
U^{(m)}(x) = W_m ^2 + W _m ^{\pr} - \beta _m
\eeq
In writing the last expression we have made use of the fact that if $v(x)$ and
$u(x)$ are isospectral, so are $V(x)$ and $U^{m}(x)$, given by 
\beq
V (x)= \{ v(x) - \beta _m \} 
\eeq
\beq
U ^{m}(x) = \{ u(x) - \beta _m \}
\eeq
For the Scarf II potential, $\beta _m $ is calculated to be 
\beq
\beta _m = \lt( p + q \rt) ^2 - 2 m \lt( p + q \rt) + m^2 \label{beta}
\eeq
Thus this approach yields new interesting potentials, with eigenfunctions
for this particular case being given by (see eq.(\ref{wavepartner})
\beq
\phi _n (x) = \lt( \f{P_m ~P_n ^{\pr}~-~ P_m ^{\pr}~P_n }{P_m } \rt) \psi _0
(x) \label{wvfnpartner}
\eeq
In the above $P_n$ stands for $P_n ^{\a, ~\b} (i~sinh~x) $ and 
$P_n ^{\pr}$ denotes differentiation of $P_n ^{\a, ~\b} (i~sinh~x) $ 
with respect to $x$. Since $P_n ^{\a,~\b}$ is well 
defined on the entire real line, so also is $\phi _n (x) $.

\noindent
Let us analyse three low lying cases $m=0,1,2$.

\vs{.5cm}

\noindent
For $m=0$
\beq
U ^{(0)} (x) 
=  - \lt \{ 2 \lt( p^2 + q^2 \rt) - (p+q) \rt \} sech ^2 x - i(p-q) 
\lt [ 2 ( p+q ) - 1 \rt ] ~ sech ~x ~ tanh ~x \label{partner0}
\eeq
with eigenenergies  
\beq
E_n = - \lt( n + 1 - p - q \rt) ^2, ~~~~~~~~~~ n = 0, 1, 2, 3, \cdots
\eeq
and the ground state 
\beq
\phi _0 = N_0 \lt( \f{1 - i~ sinh ~x }{2} \rt) ^{- \lt( p - \f{1}{2} \rt)}
\lt( \f{1 + i~ sinh ~x }{2} \rt) ^{- \lt( q - \f{1}{2} \rt)}
\eeq
Thus for $m=0$, the 
partners belong to the family of the so-called {\it satellite } 
potentials. (\ref{partner0}) is also a Scarf II potential, 
with a different set of
parameters, and shares all the energies of (\ref{scarf2}) 
except for the ground state of $V(x)$,
which is missing in (\ref{partner0}). So, 
this is analogous to the Hermitian case.

\vs{.5cm}

\noindent
$m=1$ gives the first of the non-trivial potentials. 
$$ U ^{(1)} (x) 
= - \lt \{ 2 \lt( p^2 + q^2 \rt) - (p+q) \rt \} sech ^2 x - i(p-q) 
\lt[ 2 ( p+q ) - 1 \rt]~ sech ~x ~ tanh ~x $$
\beq
+ 2 \lt( \f{f_1 ^{\pr}}{f_1} \rt) ^2 - \f{2(p-q)}{f_1} - 2 \label{partner1}
\eeq
with 
\beq
\ba {lcl}
\displaystyle f_1 (x) 
&=&  \displaystyle F \lt( -1, -p-q- \lam; 2p + \f{3}{2}; 
z \rt) \\
&=&  \displaystyle \lt \{ -(p-q) +\f{i}{2} \lt( -2p -2q +1 \rt) sinh ~x \rt \}
\ea
\eeq
The ground state of the partner (\ref{partner1}) is obtained from 
(\ref{wavepartner}) as
\beq
\phi _0 = \f{2i \lt(\f{1}{2} - p - q \rt)}{ (-p-q) + i \lt(\f{1}{2} - p - q
\rt) ~sinh ~x } ~ z^{-(p- \f{1}{2})} 
~ (z^*)^{-(q- \f{1}{2})}  
\eeq
with energy
\beq
E_0 = - \lt( - p - q \rt) ^2
\eeq
It can be shown that the state corresponding to $\psi _1$ is excluded from the spectrum as it turns out
to be non-normalizable. All other states share
identical energies with the original potential (\ref{scarf2}).
The excited states are obtained from (\ref{wavepartner}) as 
\beq
\phi_n = \lt( \f{P_1 ~P_{n+2} ^{\pr}~-~ P_1 ^{\pr}
~P_{n+2} }{P_1 } \rt) \psi _0 \label{excited}
\eeq
with energies
\beq
E_n = - \lt( n+2 - p - q \rt) ^2, n=0, 1, 2, 3, \cdots 
\eeq
As $P_n ^{\a,~\b} (i~sinh~x)$ has no zeroes on the real line, 
so the eigenfunctions are
well defined. It is easily observed from (\ref{excited}) that $\phi _n$ are 
normalizable. Moreover, the potential $U^{(1)} (x)$ so constructed has no
singularity on the real line, and hence is defined on the entire domain $( -
\infty, ~ + \infty )$. Also, for real values of the parameters $p$ and $q$
(corresponding to real energies) the new potential, too,  is invariant under 
${\cal{PT}}$ transformation.

\vs{.5cm}

\noindent
In an analogous way, the isospectral partner for $m=2$ is found to be
\beq
\ba {lcl}
\displaystyle  U ^{(2)} (x) 
&=& \displaystyle - \lt \{ 2 \lt( p^2 + q^2 \rt) - (p+q) \rt \} sech ^2 x - i(p-q) 
\lt [ 2 ( p+q ) - 1 \rt ] sech ~x ~~ tanh ~x \\
&+& \displaystyle 2 \lt( \f{ f_2 ^{\pr} }{f_2} \rt) ^2 + \f{ \sigma - 6(p-q) i~ sinh ~x
}{f_2} - 8  \label{partner2}
\ea
\eeq
with 
\beq
\displaystyle f_2 (x) 
= \displaystyle F \lt( -2, -p-q- \lam; 2p + \f{3}{2}; 
z \rt)
\eeq
\beq
\sigma = \f{ -2p-2q+2}{\lt(-2p+\f{1}{2}\rt) \lt( -2p + \f{3}{2} \rt) } 
\lt \{ 2 \lt( p - q \rt) ^2 - \f{\lt( -2q + \f{3}{2} \rt) \lt( -2p + \f{3}{2}
\rt) }{\lt( -2p -2q +2 \rt) } + \f{1}{2} \lt(3 + 2p + 2q \rt) \lt( 3 - 2p -2q
\rt) \rt \} 
\eeq 
For a visual representation as well as for comparison, we have 
plotted the real and imaginary parts of the potentials 
$V~,~U^{(i)}~~(i=0,1,2)$. In fig 1, we have plotted the real parts of $V~,~U^{(i)}~~(i=0,1,2)$ for the parameter values $V_1=24, V_2=18$ while in fig 2, we have plotted the imaginary parts of the same potentials for the same values of $V_1$ and $V_2$.

Since the Scarf II potential is always
${\cal{PT}}$ symmetric, so are its partners $U ^{(m)} (x)$, for real values
of the parameters $p$ and $q$ ( $ V_1 + 1/4 \geq |V_2| $ ), 
as the functions $f_m (x)$ remain invariant under
${\cal{PT}}$. Moreover, all the new potentials so constructed are 
defined over the entire domain $(- \infty, + \infty)$, 
admit real bound state spectrum, and possess all the energies of the
original potential except for the $m^{th}$ state, if one starts with the
$m^{th}$ order eigenfunction.
Though $m=0$ gives the usual shape-invariant form, highly non-trivial,
non-shape-invariant potentials are obtained for non-zero $m$.

\vs{1cm}

\section{Pseudo supersymmetry and intertwining} 

It is well known that in the case of Hermitian quantum mechanics, 
Darboux transformation and supersymmetric quantum mechanics are 
equivalent. Although this is not so in the case of non Hermitian 
quantum mechanics, Darboux transformation can still be implemented 
in terms of intertwining operators. To see this we consider the 
intertwining operators $A$ and $B$ :
\beq
A = \frac{d}{dx} + W_m
\eeq
\beq
B = - \frac{d}{dx} + W_m
\eeq
where $W_m$ is defined by (\ref{superpotential}), 
then the partner Hamiltonians 
\beq
H_{\pm} = - \f{d^2}{dx^2} + W_m ^2 \pm W_m ^{\pr}
\eeq
can be written as 
$H _- = BA $ and $H_+ = AB $, where
\beq
H_- = \lt( - \frac{d^2}{dx^2} + v(x) \rt) 
\eeq
\beq
H_+ = \lt( - \frac{d^2}{dx^2} + u(x) \rt) 
\eeq
Evidently, if $ \psi _n $ is an eigenfunction of $H_-$ with
energy eigenvalue $E_n ^-$, 
then $ \phi _n  = A \psi _n  $ is also an 
eigenfunction of $H_+$ with the same eigenvalue $E_n ^-$, 
escept for $ n = m $, since in this case $ A \psi _m = 0 $. 
\beq
H_+ A \psi _n = (AB)A \psi _n  = A(H_- \psi _n ) = E_n ^- (A \psi _n )
\eeq
For Hermitian Hamiltonians, $A$ and $B$ are mutually adjoint operators 
$( B = A^{\dagger} )$, 
giving the well known results of supersymmetry,
viz., $ A H_- ~=~ H_+ A $ or  $ H_- A^{\dagger} ~=~ A^{\dagger}  H_+ $.

To extend the idea of supersymmetry to include non Hermitian hamiltonians, 
we assume the existence of a linear, 
invertible, Hermitian operator $ \eta $, such that \cite{mostafa}
\beq
B ~=~ A^{\#} ~=~ \eta ^{-1}  A^{\dagger} \eta \label{pseudo}
\eeq
This allows one to rewrite the partner Hamiltonians as 
\beq
H_+ ~=~ B^{\#} B \ \ \ \ \ \ \ \ \ \   H_- ~=~ B B^{\#} 
\eeq
so that 
\beq
B H_+ = H_- B \ \ \ \ \ \ \ \ \ \ H_+ B^{\#} = B^{\#} H_- \label{def}
\eeq
From (\ref{def}) it is clear that $B$ maps eigenfunctions of $H_+$ to those of $H_-$, and $A(=B^\#)$ does the converse. Thus the mutually adjoint operators $A$ and $A^{\dagger}$
of conventional supersymmetric quantum
mechanics are replaced by their pseudo supersymmetric counterparts 
$A$ and $B$ when the potential is  non Hermitian.
However, we would like to point out that the choice of $\eta$ is not unique.

\noindent
To determine a form of $\eta$, let us note that a simple representation is given by \cite{mostafa}:
\beq
\displaystyle {\eta = {\cal P} \ \ , \ \ {\cal P} f(x) = f(-x)} \label{parity}
\eeq
It follows that for real potentials, (\ref{parity}) leads to $B = A^{\dagger}$, thus reproducing the conventional result of supersymmetry. We note that the above results are quite general since they do not depend on a specific $W_m$. We thus conclude that in the complex case the hamiltonians $H_{\pm}$ are pseudo supersymmetric partners of each other. Finally to cast the above results in a formal pseudo supersymmetric form let us define the pseudo supercharges $Q$ and $Q^{\#}$ in the following way :
\beq
Q=\left(\ba{cc}0 & A\\
0 & 0 \ea \right)~~,~~ \qquad Q^\# = \eta^{-1} Q^\dagger \eta ~=~ \left(\ba{cc}0 & 0\\
B & 0 \ea \right)
\eeq
Thus the pseudo supercharges $Q$ and $Q^{\#}$ are nilpotent and they satisfy the following closed algebra :
\beq
H = \{Q,Q^{\#}\} = \left(\ba{cc}H_+ & 0\\
0 & H_- \ea \right)~  = \left(\ba{cc}AB & 0\\
0 & BA \ea \right)~~,~~[Q,H]=[Q^{\#},H]=0~~\label{algebra}
\eeq
We thus conclude that in the non Hermitian case the hamiltonians obtained by intertwining are pseudo supersymmetric partners.

\section{A model with spontaneously broken ${\cal{PT}}$ symmetry}
As mentioned earlier, there are two cases for the ${\cal{PT}}$ symmetric 
Scarf II potential, depending on the relative strengths of $V_1$ and $V_2$, 
viz., 

\vspace{.25cm}

\noindent
(i) ~~ $\vert V_2 \vert \leq V_1 + \frac{1}{4} $ : \\
In this case, where the spectrum is real and discrete, each state is a
singlet. This has already been discussed in Section 3.

\vspace{.25cm}

\noindent
(ii) ~ $\vert V_2 \vert > V_1 + \frac{1}{4} $ : \\
In this case ${\cal{PT}}$ symmetry is  spontaneously broken.
We can choose $p$ to be real, taking a single value with only the 
positive sign in (\ref{p}) while $q$ can take either of the following values 
\begin{equation}
q^{\pm} = - \frac{1}{4} \pm i \frac{s}{2}
\end{equation}
where
\begin{equation}
s = \sqrt{ V_2 - V_1 - \frac{1}{4} }
\end{equation}
giving rise to complex conjugate pairs of energies
\begin{equation}
\ba{lcl}
E_n ^{~ \pm} &=& -\mu_n^2\pm i\mu_ns~, \ \ \ \ \ 
n = 0, 1, 2, \cdots\\
\mu_n &=& n-p+\f{1}{4}
\ea
\end{equation}
This case makes quite an interesting study and we shall investigate it
further. 
Though the original potential (14) is still ${\cal{PT}}$ 
invariant, the partners are no longer so. We note that 
while the original potential $V(x)$ does not depend explicitly on 
the parameters $p$ and $q$, the partner potential is explicitly
dependent on these parameters. As a consequence there are 
two partner potentials corresponding to $V(x)$ 
(this is due to the fact that in this case both the 
values of $q$ are allowed). 
If the Darboux transformation is carried out by the $m^{th}$ eigenstate
$ \psi _m ^{-} $ ($ \psi _m ^{+} $) 
of the original potential, then straightforward calculations
show that the corresponding state will be missing in the partner. 
Furthermore, the entire positive (negative) sector 
$ E_n ^{+} = - \left( \mu _n ^2 - \frac{s^2}{4} \right) + i \mu _n s $
~~~($ E_n ^{-} = - \left( \mu _n ^2 - \frac{s^2}{4} \right) - i \mu _n s $)~
will be absent in the partner. 
Thus the spectrum of the partner potential 
is quite different from that of the Scarf II potential, as 
the former has only singlet complex energies.
So, if one starts with the ground state eigenfunction $ \psi _0 ^- $, 
then the partner potential (\ref{partner0}) is of the form
\beq
U_-^0(x) = -\{(2p^2-p-\f{s^2}{2}+\f{3}{8})+is\}~sech^2x~-~\{i(2p^2-p+\f{s^2}{2}-\f{3}{8})+s\}~sechx~tanhx
\eeq
while the wave functions and the corresponding energy values are given by
\begin{equation}
\phi_n^-(x) = \f{ab}{c}z^{-p}(z^*)^{-\f{1}{4}+\f{is}{2}} F(a+1,b+1;c+1;z)~~,~~E_n ^{~-} =-\mu_{n+1}^2-i\mu_{n+1}s ~, 
\ \ \ \ \ n = 0, 1, 2, \cdots
\end{equation}
where
\beq
a=-(n+1)~,~b=(n+\f{3}{2}-2p)+is~,~c=-2p+\f{1}{2}
\eeq

\section{Discussion}
In this article we have explored the idea of applying Darboux transformation to non Hermitian quantum mechanical systems. In particular we have obtained a series of new exactly solvable ${\cal{PT}}$ symmetric potentials with real bound state spectra. The symmetry aspect of the potentials has also been investigated and it has been shown that they are pseudo supersymmetric. On the other hand when Darboux transformation is applied to a system with spontaneously broken ${\cal{PT}}$ symmetry the resulting potential admits a single tower of energy. We would like to point out that this is not a characteristic feature of only models with spontaneously broken ${\cal{PT}}$ symmetry but in general the same result would be obtained whenever the original model posseses doublet of states \cite{znojil1}.

\section*{Acknowledgment}
The authors would like to thank the referees for suggesting improvements. The authors also thank A. Mostafazadeh for pointing out an error. One of the authors (A.S.) thanks the Council of Scientific \& Industrial Research,
India, for granting her Senior Research Associateship. 

\pb


\newpage

\noindent
{\bf Figure Captions}

\vs{1cm}

\noindent
Fig. 1 : Graph of real parts of $V$ (solid),
~$U^{(0)}$~(dotted),~$U^{(1)}$~ (small dash),~$U^{(2)}$~ (large dash).

\vs{1cm}

\noindent
Fig. 2 : Graph of imaginary parts of $V$ (solid),
~$U^{(0)}$~(dotted),~$U^{(1)}$~ (small dash),~$U^{(2)}$~ (large dash). 

\end{document}